# Community-Based Data Integration of Course and Job Data in Support of Personalized Career-Education Recommendations

**Guoqing Zhu**
School of Maritime Economics and Management
Dalian Maritime University
Dalian, 116026, China
zhuguoqing@dlmu.edu.cn

**Naga Anjaneyulu Kopalle**
Luddy School of Informatics, Computing and Engineering
Indiana University Bloomington
Bloomington, IN 47408, USA
nakopa@indiana.edu

**Yongzhen Wang**
Luddy School of Informatics, Computing and Engineering
Indiana University Bloomington
Bloomington, IN 47408, USA
wang11@indiana.edu

**Xiaozhong Liu**
Luddy School of Informatics, Computing and Engineering
Indiana University Bloomington
Bloomington, IN 47408, USA
liu237@indiana.edu

**Kemi Jona**
Northeastern University
Boston, MA 02115, USA
k.jona@northeastern.edu

**Katy Börner**
Luddy School of Informatics, Computing and Engineering
Indiana University Bloomington
Bloomington, IN 47408, USA
katy@indiana.edu

## ABSTRACT

How does your education impact your professional career? Ideally, the courses you take help you identify, get hired for, and perform the job you always wanted. However, not all courses provide skills that transfer to existing and future jobs; skill terms used in course descriptions might be different from those listed in job advertisements; and there might exist a considerable skill gap between what is taught in courses and what is needed for a job. In this study, we propose a novel method to integrate extensive course description and job advertisement data by leveraging heterogeneous data integration and community detection. The innovative heterogeneous graph approach along with identified skill communities enables cross-domain information recommendation, e.g., given an educational profile, job recommendations can be provided together with suggestions on education opportunities for re- and upskilling in support of lifelong learning[1].

## 1 Introduction

Lifelong learning, the pursuit of knowledge for either personal or professional reasons, enhances employees'/professionals' competitiveness and career satisfaction. Students often explore various kinds of educational opportunities to reach their career goals. Generally, the education system should serve the career ecosystem, and skill discrepancies between research, education, and jobs should be minimized (Börner et al., 2018).

[1] This work was partially supported by the National Science Foundation under award 1936656. Any opinions, findings, and conclusions or recommendations expressed in this material are those of the author(s) and do not necessarily reflect the views of the NSF.

From an information recommendation perspective, although existing job recommendation systems meet some of the needs of students, their impact is limited. Generally, job/career and course/education recommendation systems implement the same multi-task learning framework: Given a user's career, educational, or hybrid information needs, recommend various types of jobs/courses simultaneously. Notwithstanding a rich body of literature on course recommendation and job recommendation, few studies consider both of them simultaneously (Li et al., 2017).

In this study, we propose a novel method to integrate job/career and course/education data by employing heterogeneous graph indexation. As shown in Figure 1, two different sources of data on education and career development are integrated into a heterogeneous graph, with skills serving as the bridge. Because job skills and course skills are different, skill communities computed by the Infomap algorithm, help connect jobs and courses data. Given a properly indexed heterogeneous graph, we can then apply a random walk algorithm to generate personalized suggestions for courses and jobs by considering future career goals or planned educational experiences.

To showcase that students are able to benefit from this graph-based data integration, we conducted preliminary experiments, using the course data from Indiana University and IT industry job postings. Results demonstrate that the proposed approach can effectively provide course recommendations in light of given career goals. The proposed method can be generalized to different education/career environments.

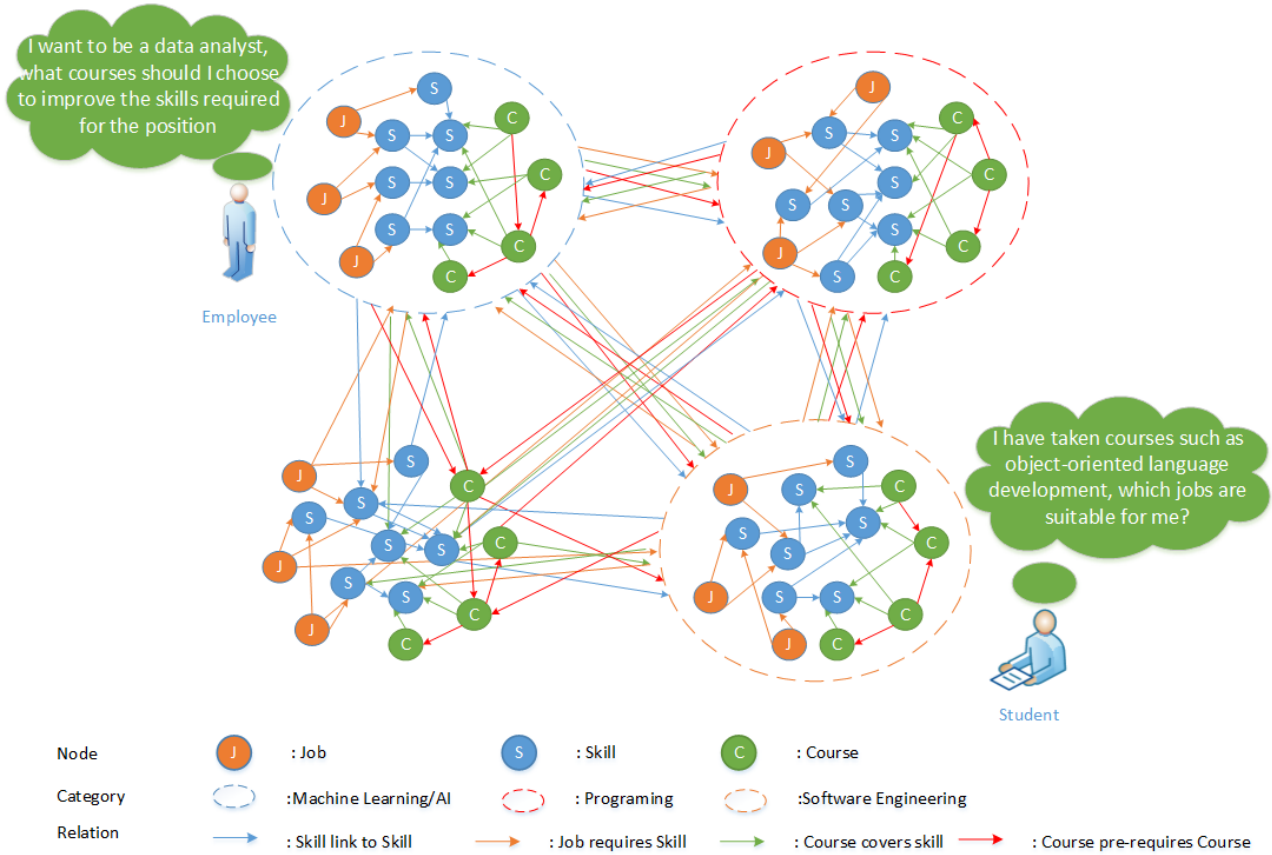

Figure 1: **Heterogeneous Graph Index Schema.** The orange circles represent the job nodes, the blue circles represent the skill nodes, and the green circles represent the course nodes. The directed orange edges represent the job to skill relations, the directed blue edges represent the skill that was required by a job to a skill that was covered by a course via the linked relation, the directed green edges represent the course to skill via the covered relation, the directed red edges represent the course to the course via the pre-required relation. The dotted circle indicates the skill community to which the job and course belong to.

## 2 Literature Review

Recommender systems have been broadly applied in the context of course planning. For most of these studies, courses were recommended to target users based on other users' feedback, overall user performance or similarities between course materials (Li et al., 2017; Wang et al., 2017). For instance, Nguyen et al. (2018) applied sequential rule mining for pairs of courses and grades and recommend the course with the best performance. Generally, few course recommendation systems consider users' future career goals or target jobs (Ma and Ye, 2018). Many graph-based course recommender systems have been developed. For instance, Bridges et al. (2018) made personalized suggestions about which course should he/she enroll based on a directed graph that gathered students' grades and enrollment history. However, existing graph-based course recommendation research focuses on the education domain only. To the best of our knowledge, this study is one of the first investigations of graph-based cross-domain recommendation that leverages massive education and career data via community-based data fusion.

In recent decades, similar to course recommendation systems, job recommendation systems have generated tremendous interest in the research community. Some researchers have studied job recommendations from the perspective of career paths (Patel et al., 2017). Some approaches used social networks to generate job recommendations. Shalaby et al. (2017) proposed a graph-based approach that uses the relationship between user-work interaction and job posting content for real-time job recommendations. Broadly speaking, prior research work on job recommendations mainly focuses on information about the user's professional experience, without considering the user's educational history. Besides, although some studies use graph-based methods, they only focus on a single career domain.

This study goes beyond the existing work by presenting a novel graph-based approach that recommends rank-ordered courses or jobs for a student (or junior employee) by considering his/her education/career history and leveraging a heterogeneous graph that integrates education and career data.

## 3  Research Methods

In this section, we discuss the proposed method in detail, which includes: collecting career data and education data (Section 3.1), integrating two datasets using skills communities computed using Infomap plus heterogeneous graph and indexing (3.2), ranking courses (as a case study) via a graph-enabled cross-domain ranking function that uses a random walk algorithm (3.3), and running a preliminary user study (3.4).

### 3.1  Data Collection

The dataset collected in this project includes two types of data:

**Courses/Education data** was gathered from the course enrollment logs of the Luddy School of Informatics, Computing, and Engineering (SICE), Indiana University Bloomington (IUB), covering four academic years from 2016 to 2019. This data consists of 7,824 students, 371 courses, and 188,881 records of course enrollment from 5 departments over 16 academic semesters. Original course data does not have information on the skills taught by a course. In order to interlink the courses and skills, we extracted 957 MOOCs[2] and 1011 related skills in the field of computer science, informatics, information and library science, intelligent system engineering, and statistics via automated web scraping techniques. By leveraging the greedy match algorithm, all the IUB-SICE courses are projected to related skills from MOOCs. At the end, each course in the educational data set has four features (i.e., the course ID, the course name, the course description, and all related skills). A total of 266 university courses and 376 skills are included in the educational dataset.

**Job/Career dataset** was compiled from Careerbuilder[3] job advertisements downloaded in December 2019. Popular IT job titles[4] were used as the search query, redundant jobs were removed, and the final data comprises a total of 20,000 jobs and the 1,611 skills associated with them. The job advertisements were analyzed and five features were extracted: the job ID, the job title, the company, the location, and the list of required skills.

### 3.2  Heterogeneous Graph-based Data Indexation and Skill Community Detection

The main bridge for integrating career and education data is a set of skills required by a profession and a set of skills covered by each course (Li et al., 2017). However, skills listed in courses and skills listed in jobs differ–they utilize different vocabularies. In our dataset, we only identified 79 overlapping skills across job and course data. In order to address this challenge, we utilize the Infomap algorithm (Rosvall and Bergstrom, 2008) to detect skill communities in the target graph $M$. The Infomap method works as follows: simulates a random walker wandering on the graph for $m$ steps and indexes his random walk path via a two-level codebook (a global index codebook and each community having a code book). The goal is to generate a community partition with the minimum random walk description length,

---

[2] https://www.coursera.org/
[3] https://www.careerbuilder.com/
[4] https://money.usnews.com/money/careers/slideshows/discover-the-best-technology-jobs

which is calculated as follows:

$$L(\pi) = \sum_i^m q^i H(Q) + \sum_{i=1}^m p^i H(\mathcal{P}^i) \quad (1)$$

Where $L(\pi)$ is the description length for a random walker under current community partition $\pi$. $q^i$ and $p^i$ are the jumping rates between and within the $i_{th}$ community in each step. $H(Q)$ is the frequency-weighted average length of codewords in the global index codebook and $H(\mathcal{P}^i)$ is the frequency-weighted average length of codewords in the $i_{th}$ community codebook.

We first create a career graph and an education graph individually. By using Infomap, all the skill nodes are grouped into communities (e.g., *programming community* and *AI community*). Then, the most similar career and education communities (by counting the overlapping skills) are merged along with the associated jobs and courses (see Figure 1).

By using this method, all the education and career data (skills, courses and jobs) are integrated into the same heterogeneous graph $G = (V, E)$. In this graph, we have defined a node type mapping function $\tau: V \rightarrow O$ and an edge type mapping function $\phi: E \rightarrow R$, where each node $v \in V$ belongs to one particular variable $\tau(v) \in O$, and each edge $e \in E$ belongs to one particular relation $\phi(e) \in R$. If two links belong to the same relation type, the two links share the same starting object type and the ending object type. The nodes and relations are described in Table 1.

Table 1: Nodes and Relations in the constructed heterogeneous graph

| Nodes and Relations | Description |
|---|---|
| $C$ | The course nodes |
| $J$ | The job nodes |
| $S$ | The skill nodes |
| $C \xrightarrow{p} C$ | The course to course edge via the pre-required relation |
| $C \xrightarrow{c} S$ | The course to skill edge via the covered relation |
| $J \xrightarrow{r} S$ | The job to skill edge via the required relation |
| $S \xrightarrow{l} S$ | Skill to skill edge (skill-skill text similarity within each community based on BM25). |

For any node on the graph, the sum of the same type of outgoing links equals 1. For instance, the weight of the link from $C_i$ to $C_j$ is defined as $w(C_i \xrightarrow{p} C_j) = \frac{d(C_i \xrightarrow{p} C_j)}{d(C_i \xrightarrow{p} C)}$, where $d(C_i \xrightarrow{p} C_j)$ is the number of students who enrolled for course $C_j$ before enrolling for course $C_i$, and $d(C_i \xrightarrow{p} C)$ is the total number of students who enrolled for any course before enrolling course $C_i$. The weight of $C_i \xrightarrow{c} S_j$ is defined as $w(C_i \xrightarrow{c} S_j) = \frac{1}{d(C_i \xrightarrow{c} S)}$, where $d(C_i \xrightarrow{c} S)$ is the total number of skills covered by course $C_i$. The weight of $J_i \xrightarrow{r} S_j$ is defined as $w(J_i \xrightarrow{r} S_j) = \frac{d(J_j \xrightarrow{r} S_j)}{d(J_i \xrightarrow{r} S)}$, where $d(J_i \xrightarrow{r} S_j)$ is the number of job $J_i$ that required skill $S_j$, and $d(J_i \xrightarrow{r} S)$ is the total number of job $J_i$ that required any skill. The relation $S \xrightarrow{l} S$ is the key to internally connecting the entire heterogeneous graph. The weighted of $S \xrightarrow{l} S$ is normalized similarity score (via BM25) between skills within each community. Because of community restriction, the noisy similar skill pairs won't pollute the graph accuracy.

The complete network graph has a total 22,253 nodes and 95,712 edges. There are 20,000 jobs, 266 university courses, 1,987 course and job skills, and the numbers of the various relations are: 73,560 $J \xrightarrow{r} S$, 11,155 $C \xrightarrow{p} C$, 641 $C \xrightarrow{g} S$, 10,356 $S \xrightarrow{sim} S$ respectively.

### 3.3 Heterogeneous Graph-enabled Cross-Domain Recommendation with Community Restriction

In this section, we apply a graph-enabled cross-domain ranking approach to make education/career recommendations. For each query node in the graph, we retrieve target candidate nodes and make suggestions based on their ranking scores. The ranking score of each candidate node comes from the meta-path-based ranking function (Liu et al., 2014) along with the community structure. On the heterogeneous graph, the meta-path defines the connection between query nodes and result nodes. For the same recommendation task (for example, recommending courses to different types of users), there are usually multiple meta-paths on the heterogeneous graph. Besides, when we change the type of query node and recommended node, this method can be generalized to other recommendation tasks, e.g., recommending a job to a student or professional.

To quantify the ranking score of a candidate's nodes following the meta-path, a random walk-based measure is proposed (Liu et al., 2014). It can be represented by:

$$s(v_i^{(1)} \to v_j^{(l+1)}) = \sum_{t=v_i^{(1)} \to v_j^{(l+1)}} RW(t) \qquad (2)$$

Where $v_i$ represent the seed node, and $v_j$ is for a candidate's queried node. Where $t$ is a tour from $v_i$ to $v_j$. Suppose $t = (v_{i1}^{(1)}, v_{i2}^{(2)}, ..., v_{il+1}^{(l+1)})$. The random walk probability is then $RW(t) = \prod_j w(v_{ij}^{(j)}, v_{ij+1}^{(j+1)})$, where $w(v_{ij}^{(j)}, v_{ij+1}^{(j+1)})$ is the weight of edge $v_{ij}^{(j)} \to v_{ij}^{(j+1)}$.

As a proof of concept, we conducted a course recommendation user study and defined different meta-paths to recommend courses for three scenarios:

**Scenario 1:** A first-year undergraduate/graduate student has a career goal (job node), and he/she is looking for education suggestions (courses nodes) to achieve this career goal.

For this scenario, the input is the student's career goal (job node), and the output is a set of recommended candidate courses (nodes). The corresponding meta-path function is:

$$\mathbb{C}_J || J \xrightarrow{r} S \xrightarrow{l} S \xleftarrow{c} C_?$$

$$\mathbb{C}_{\not J} || C_? \xleftarrow{p} C_?^p$$

Where $J$ is the query job node, and $C_?$ is the candidate course node. This path function, walking through the relationship between the job and courses via skills, the candidate courses related to the career goal are retrieved. Note that the first function only performs on the target community $\mathbb{C}_J$ that job $J$ belongs, and the second function retrieves all required courses $C_?^p$ from communities $\mathbb{C}_{\not J}$ that the job $J$ does not belong to.

**Scenario 2:** A student already took some courses $C^p$, and he/she is looking for new courses $C_?$ to achieve his/her career goal $J$. This is similar to the Scenario 1 but we add another function: $\mathbb{C}_{\not J} || C^p \xrightarrow{c} S \xleftarrow{c} C_?$ where $C^p$ has additional chance to help locate relevant courses $C_?$.

**Scenario 3:** An employee/professional has a current job J , and he/she is looking for career acceleration. That is, he/she is looking for a course that helps to upskill. The ranking function could be $J \xrightarrow{r} S \xrightarrow{l} S \xleftarrow{c} C \xrightarrow{p} C_?$ Because user's information need is to upskill, the last step will be $C \xrightarrow{p} C_?$; the foundation course (like programming) can walk to a more advanced course (like machine learning).

Unlike prior studies on this track, community restriction is critical for the proposed random walk function, which can be useful to reduce noise and enhance the recommendation accuracy.

### 3.4 Preliminary Experimental Result

A preliminary experiment was run for Scenario 1, where two graduate students at IUB were asked to use the course recommendation system by leveraging the proposed ranking algorithm. Each student entered 5 text queries (e.g., *"Database Administrator"*, *"Java Developer"*, and *"Data Scientist"*) and rated each recommended course as *"useful"* or *"not useful"*. MAP (Mean Average Precision) or Precision was used as the evaluation metrics. For MAP case, binary judgment is provided for each candidate course. We also evaluate the ranking performance with a given cut-off rank, considering only the topmost candidate courses returned by the experiment. In Table 2, we report the performance of different recommendation functions (overall and top-ranked education opportunities performance). In the experiment, two baseline methods (Vector Space model and Probabilistic Model (Truyen et al., 2014) are employed for comparison.

Based on this table, the proposed method outperforms the baselines for most of the evaluation metrics except for precision. The result shows that the proposed method is promising for cross-domain recommendation while students/professionals can potentially benefit from it. Note that, in this experiment, we did not utilize sophisticated graph ranking models and learning to rank algorithms. Prior studies (Liu et al., 2014) showed that these methods could further enhance the recommendation performance. We plan to explore more sophisticated graph ranking models in future investigations.

Table 2: Preliminary Result of Course Recommendation for Different Ranking Features

|  | Precision | MAP | MAP@5 | Precision@10 | MAP@10 |
|---|---|---|---|---|---|
| Vector Space | 0.41 | 0.50 | 0.50 | 0.48 | 0.51 |
| Probabilistic Model | 0.38 | 0.55 | 0.55 | 0.44 | 0.59 |
| Graph-Enabled | 0.39 | 0.57 | 0.63 | 0.48 | 0.62 |

## 4 Conclusions

In this study, we proposed a novel method for career and education data integration and indexation by using a heterogeneous graph and community detection. The generated graph enables different information retrieval/recommendation scenarios to address various kinds of information needs from students and professionals. From a data science perspective, career data and education data need to be cross-walked, and the community detection along with data fusion provides a promising method for data integration.

The limitation of this method is that there are fewer overlapping skills across jobs and courses—only 79 of the 376 skills identified in the course data could be mapped to skill terms listed in job advertisements.

In the future, our efforts on this project will be threefold. Firstly, we would like to enhance the graph quality by adding novel nodes and edges, e.g., company, location, and enrollment information. Secondly, we would like to investigate a sophisticated graph ranking algorithm to enhance the recommendation performance. Last but not least, we plan to conduct more comprehensive user evaluations employing both professionals and students. User feedback will be used for model training and user interface optimization.

## References

Börner, K., Scrivner, O., Gallant, M., Ma, S., Liu, X., Chewning, K., Wu, L., and Evans, J. A. (2018). Skill discrepancies between research, education, and jobs reveal the critical need to supply soft skills for the data economy. *Proceedings of the National Academy of Sciences*, 115(50):12630–12637.

Bridges, C., Jared, J., Weissmann, J., Montanez-Garay, A., Spencer, J., and Brinton, C. G. (2018). Course recommendation as graphical analysis. *2018 52nd Annual Conference on Information Sciences and Systems, CISS 2018*, pages 1–6.

Li, N., Naren, S., Gao, Z., Xia, T., Börner, K., and Liu, X. (2017). Enter a job, get course recommendations. *iConference*, 2:118–122.

Liu, X., Yu, Y., Guo, C., and Sun, Y. (2014). Meta-path-based ranking with pseudo relevance feedback on heterogeneous graph for citation recommendation. In *Proceedings of the 23rd ACM international conference on conference on information and knowledge management*, pages 121–130.

Ma, X. and Ye, L. (2018). Career goal-based e-learning recommendation using enhanced collaborative filtering and prefixspan. *International Journal of Mobile and Blended Learning (IJMBL)*, 10(3):23–37.

Nguyen, H.-Q., Pham, T.-T., Vo, V., Vo, B., and Quan, T.-T. (2018). The predictive modeling for learning student results based on sequential rules. *Int. J. Innov. Comput. Inf. Control*, 14(6):2129–2140.

Patel, B., Kakuste, V., and Eirinaki, M. (2017). Capar: a career path recommendation framework. In *2017 IEEE Third International Conference on Big Data Computing Service and Applications (BigDataService)*, pages 23–30. IEEE.

Rosvall, M. and Bergstrom, C. T. (2008). Maps of random walks on complex networks reveal community structure. *Proceedings of the National Academy of Sciences*, 105(4):1118–1123.

Shalaby, W., AlAila, B., Korayem, M., Pournajaf, L., AlJadda, K., Quinn, S., and Zadrozny, W. (2017). Help me find a job: A graph-based approach for job recommendation at scale. In *2017 IEEE International Conference on Big Data (Big Data)*, pages 1544–1553. IEEE.

Truyen, T. T., Phung, D. Q., and Venkatesh, S. (2014). Preference networks: Probabilistic models for recommendation systems. *arXiv preprint arXiv:1407.5764*.

Wang, Y., Liu, X., and Chen, Y. (2017). Analyzing cross-college course enrollments via contextual graph mining. *PloS one*, 12(11).